\documentclass[aps,nofootinbib,floats,superscriptaddress]{revtex4-2}

\usepackage{graphicx}
\usepackage{epstopdf}
\usepackage{amssymb}
\usepackage{amsmath,bm}
\usepackage{psfrag}
\usepackage{epsfig}
\usepackage{float}
\usepackage{bm}
\usepackage{textcomp}
\usepackage{hyperref}

\begin{document}

\title{Nanowire Breakup via a Morphological Instability Enhanced by Surface Electromigration}

\author{Mikhail Khenner\footnote{Corresponding
author. E-mail: mikhail.khenner@wku.edu.}}
\affiliation{Department of Mathematics, Western Kentucky University, Bowling Green, KY 42101, USA}
\affiliation{Applied Physics Institute, Western Kentucky University, Bowling Green, KY 42101, USA}

\begin{abstract}
\noindent

Using a recent continuum model of a single-crystal nanowire morphological evolution in the applied axial electric field, 
an axisymmetric evolution of a microscopically rough nanowire surface is computed. 
Morphological evolution results in a wire breakup into a cylindrical segments (particles). 
Breakup time and the number of particles are characterized for various levels of the radial and axial surface roughness.  
It is shown that electromigration and larger surface roughness lead to a shorter breakup time and the increased number of particles.

\end{abstract}

\date{\today}
\maketitle


\section{Introduction}
\label{Intro}

It is well-known that solid cylindrical nanowires are vulnerable to a Rayleigh-Plateau-type morphological instability, that typically results 
in a local wire thinning and the ultimate breakup (with a subsequent formation of a chain array of nanoparticles) \cite{XLL,KT,KTBECKN,KTEBCKN,LBPDVC}.  
The driving force for morphological evolution is a high-temperature surface diffusion, whereby atoms in the surface layer diffuse from the regions of 
high surface chemical potential to the regions of low potential \cite{NM2,NM1}. 
If the axial electric current is passed through a wire, the released Joule heat makes surface diffusion stronger, but more important, the current provides
the additional mass transport mechanism known as the surface electromigration \cite{Huntington,HoKwok,REW,SK}. The essence of this effect is that the 
``electron wind" sweeps surface ions in the average direction of the current via scattering.

Effects of surface electromigration\footnote{For the rest of this letter, called simply ``electromigration".} on morphological and compositional 
instabilities and evolutions of crystal surfaces were considered experimentally and theoretically 
since 1960s (see Ref. \cite{MyWire1} for references to major modeling and computational works in this area). In addition, morphological evolution of wires by surface diffusion, 
absent electromigration, received a lot of attention in the modeling and applied mathematics communities in 1980s and 1990s \cite{C,CFM,McCVoorhees,BBW,WMVD,KDMV,GM,GMM}. 
Presently, understanding electromigration-driven instabilities and breakup of wires is important for reliable and scalable manufacturing of 
nanocontacts \cite{PLAPM,VFDMSKBM,AGLCH}. Advancements in that area may enable future breakthrough applications, such as measurement of the electrical 
conductivity and surface electron states of a single organic molecule.
It is therefore very strongly surprising that a morphological 
evolution of metallic wires subjected to electromigration was not explored via basic theory and modeling, until very recently \cite{MyWire1}.
In that paper, we considered the base linear stability problem for ``electromigrated"  nanowire that is either free-standing or is pre-grown on a 
substrate, such that it makes two contact lines with the latter. 
It was demonstrated that in the applied axial electric field a wire is destabilized by a perturbations having wavelengths 
shorter than a wire circumference, which differs from the case without electromigration \cite{NM2,NM1}.
Stronger electric field results in a wider interval of unstable wavelengths, a decrease of a most dangerous unstable wavelength, and an increase of a 
maximum growth rate of the instability. Also it was noted that in a stronger electric field a wire crystallographic orientation has larger impact on stability. 


Historically, the models of a breakup (rupture) of thin solid (liquid) films were focused on exploration of a surface evolution self-similarity and a 
singularity formation at a breakup \cite{BBW,WMVD} or rupture \cite{BLS,Egg,Wit} point. While these questions are deep, fascinating, and mathematically rich, an experimentalist 
would be more interested in a more pragmatic issues that are motivated by technological applications, such as: 
What is the impact of (thermal) surface roughness on a time to breakup ? What is a number of segments into which a a wire would break,
for a wire of given length, radius, and surface roughness ? 
This letter aims to provide some answers to these questions by computing a morphology evolution partial differential equation until a wire breaks up. 
Since the initial surface roughness
is a random variable, the approach in this work rests on computing multiple realizations of a nanowire surface morphological evolution for various (quasi)random initial conditions 
and averaging the results.

\section{The model}
\label{Model}

In this section, the model is briefly presented; full details and extensive discussions are in Ref. \cite{MyWire1}. 

Considered is a free-standing cylindrical monocrystalline nanowire of the radius $R_0$. 
A constant electric field $E_0$ is applied 
along the wire axis (assumed directed along the $y$-axis), inducing electromigration; a local approximation for the surface 
electric field is adopted. 
For axisymmetric dynamics at all times, a wire surface is described by $r=F(t,y)$, where $r$ is the radial variable.
The angle $\phi$ that quantifies the slope of the surface in $yz$ plane is introduced, as well as the length scale $L$ and the timescale $\tau$:
$F_y=\tan{\phi}$, $L=R_0$, $\tau=L^4 k T/(D\gamma\Omega^2\nu)$. Here, $k T$ is the thermal energy, 
$\gamma$ the surface energy, $\Omega$ the atomic volume, and $D,\ \nu$ the surface diffusivity and the density of the adatoms, respectively. 
The dimensionless variables are defined as follows: $\tilde r =r/L$, $\tilde y = y/L$, $\tilde \kappa = \kappa L$, $\tilde F = F/L$. Here $\kappa$ is 
the mean curvature of the surface.

Dropping the tildas, the dimensionless equation describing the evolution of the wire surface reads:
\begin{equation}
F_t=\frac{1}{F} \frac{\partial}{\partial y}\left[M(\phi)\frac{F \kappa_y + E}{\left(1+F_y^2\right)^{1/2}}\right],
\label{Vn}
\end{equation}
where
\begin{equation}
\kappa=\frac{1}{F\left(1+F_y^2\right)^{1/2}}-\frac{F_{yy}}{\left(1+F_y^2\right)^{3/2}}
\label{n_and_curv}
\end{equation}
and
\begin{equation}
M(\phi)=\frac{1+S\cos^2{\left[m\left(\phi+\psi\right)\right]}}{1+S},\quad \phi=\arctan{F_y}
\label{mobility}
\end{equation}
is the anisotropic diffusional mobility of the adatoms \cite{SK}. 
In Eq. (\ref{mobility}) $S$ is the anisotropy strength, $m=1,2,$ or $3$ the number of symmetry axes, and the misorientation angle $\psi$ 
is the angle between a crystalline symmetry direction, such as [110], [100], or [111], and the average surface orientation 
(i.e., the average orientation in the $yz$ plane of the unit normal to the surface). A crystalline symmetry direction 
also determines the number of symmetry axes.
For [110] direction: $m=1$, $0\le \psi\le \pi/2$; for [100] direction: $m=2$, $0\le \psi\le \pi/4$; 
for [111] direction: $m=3$, $0\le \psi\le \pi/6$ \cite{DM}. 
When surface diffusion is isotropic ($M(\phi)=1$) 
and the electric field is off ($E=0$), Eq. (\ref{Vn}) reduces to a well-known (axisymmetric) evolution equation \cite{BBW,CFM}
\begin{equation}
F_t=\frac{1}{F} \frac{\partial}{\partial y}\left[\frac{F \kappa_y}{\left(1+F_y^2\right)^{1/2}}\right],
\label{AxiBBW}
\end{equation}
which is the mathematical expression of the adatom surface diffusion via a surface Laplacian of mean curvature in cylindrical coordinates \cite{Mullins95,CT94}.
 
The dimensionless electric field $E=Q E_0 R_0^2/(\Omega \gamma)=Q \Delta V R_0^2/(\Omega \gamma d)$,
where $Q>0$ is the effective charge of ionized atoms, $\Delta V$ applied voltage (to the front and back faces of the wire), and $d$ the wire length.
The typical values of the physical parameters are: $R_0=25$nm \cite{KTBECKN}, $Q=10^{-9}$C \cite{REW}, 
$\gamma=2500$erg$/\mbox{cm}^2$, $\Omega=10^{-22}$cm$^3$, 
and $d=1000$nm \cite{KTBECKN}. Then at $\Delta V=1$V, $E=250$. For reasons of numerical stability and reproducibility, in 
this paper the computations are executed at weaker electric fields, $0\le E \le 5$. Values of $S$, $m$, and $\psi$ will be fixed: $S=m=1$, $\psi=\pi/12$.

\section{Results}
\label{Res}

\begin{figure}[H]
\vspace{-0.2cm}
\centering
\includegraphics[width=5.0in]{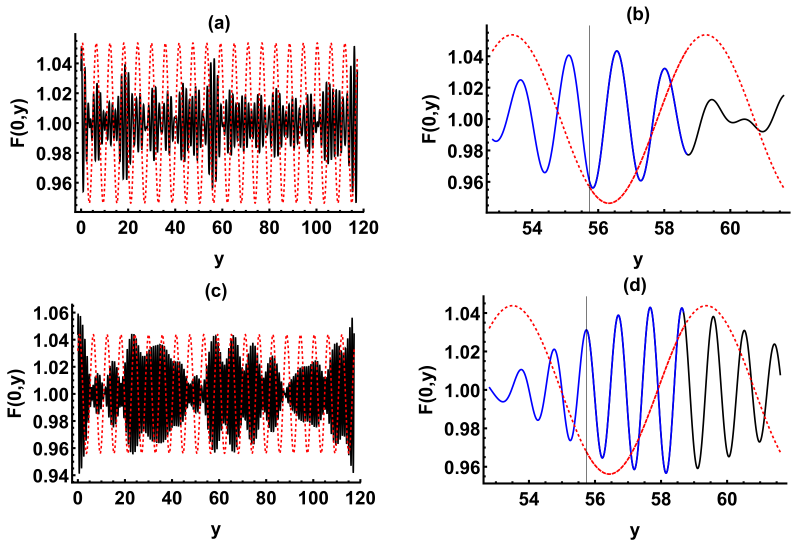}
\vspace{-0.15cm}
\caption{(a) Plot of Eq. (\ref{perturb}) at $(n_1,n_2)=(70,90)$ and $a=0.15$. For comparison, the dashed curve shows the plot of that equation at 
$(n_1,n_2)=(20,20)$; notice that the amplitude of that curve is not a constant, but is slowly varying. (b) Zoom-in view of the panel (a). Blue curve is the section of the black curve in panel (a) on the interval $[9\lambda_{max},10\lambda_{max}]$. 
Blue-and-black curve to the right of the thin vertical line is the section of the black curve in panel (a) on the interval $[9.5\lambda_{max},10.5\lambda_{max}]$. 
(c), (d): Same as (a) and (b), respectively, but $(n_1,n_2)=(110,130)$. The change in the length of the red dash in (b) and (d) is the artifact of the plotting software 
that should be disregarded.
}
\label{Fig_shapes_t=0}
\end{figure}

Initial wire shape is given by equation $F(t,y)=1$. A quasi-random perturbation of this shape is introduced to serve as the initial condition for Eq. (\ref{Vn}):
\begin{equation}
F(0,y)= 1+a \sum_{k=1}^{10} \left[\frac{1}{\mbox{RI}(10,100)} \cos{\left(\frac{k_{max}}{20}\mbox{RI}(n_1,n_2)y\right)}+
\frac{1}{\mbox{RI}(10,100)} \sin{\left(\frac{k_{max}}{20}\mbox{RI}(n_1,n_2)y\right)}\right],\quad 0\le y\le \ell
\label{perturb}
\end{equation}
where $a$ is the ``amplitude", $\ell=20\lambda_{max}=40\pi/k_{max}$ is the size of the computational domain (with $k_{max}, \lambda_{max}$ the most dangerous 
wavenumber and the most dangerous wavelength from linear stability analysis \cite{MyWire1}),  
\begin{equation}
k_{max}=\sqrt{\frac{1}{2}\left(1+E\frac{2m S \sin{2m \psi}}{2+S(1+\cos{2m \psi})}\right)},\quad \lambda_{max}=\frac{2\pi}{k_{max}},
\label{kc_axi_alpha_pi}
\end{equation}
and $\mbox{RI}(u,v)$ is a random integer in $[u,v]$ ($u$ and $v$ are integers). Notice that for each $k=1,...,10$ the random number generator is called 
four times; also, at $(n_1,n_2)=(20,20)$
the perturbation has the most dangerous wavelength $\lambda_{max}$. By varying $a$ and the integer pair $(n_1,n_2)$, the roughness of the initial wire surface is set.
For multiple realizations of Eq. (\ref{perturb}) with fixed $a$ and various pairs $(n_1,n_2)$, the root-mean-square roughness 
$R_{rms}=\sqrt{(1/\ell)\int_{0}^{\ell} F(0,y)^2 dy}$ is practically constant. 
Thus in order to increase (decrease) $R_{rms}$ it is sufficient to increase (decrease) a single parameter, $a$. Note that $R_{rms}$ is the measure of 
surface roughness in the radial ($z$) direction.  On the other hand,  simultaneously increasing 
$n_1$ and $n_2$ according to a certain protocol, 
or simultaneously decreasing $n_1$ and $n_2$ (at fixed $a$) allows to ``pack" more or fewer perturbation ``wavelengths" into a single most unstable 
wavelength $\lambda_{max}$. This way the axial roughness is controlled. For illustration, Fig. \ref{Fig_shapes_t=0} shows plots of $F(0,y)$ for $(n_1,n_2)=(70,90)$ and $(n_1,n_2)=(110,130)$. 
Note that these are only two realizations of the initial condition shown as example - for a chosen $a$ and $(n_1,n_2)$ pair I use Eq. (\ref{Vn}) 
to compute evolution of multiple realizations of the initial condition. 
$R_{rms}=0.017$ for the cases shown in Figures \ref{Fig_shapes_t=0}(a,c), which translates to the dimensional value 0.425 nm. 
For comparison, covalent and metallic diameters of a Pt atom are 0.22 and 0.28 nm. (Pt is chosen as the example, since this metal is commonly used 
to grow nanowires and nanoclusters.)
Thus $R_{rms}$ value around 0.01-0.02 corresponds to the atomic-scale radial roughness that is naturally present on material surfaces after their 
synthesis or growth. 
On the other hand, the horizontal distance between neighbor extrema of the initial profile is much larger than the atomic scale, 
as seen in Figures \ref{Fig_shapes_t=0}(b,d); in fact, in Fig. \ref{Fig_shapes_t=0}(d) it is two orders of magnitude larger. 
A larger axial scale for the roughness is expected for crystalline surfaces, as they 
are composed of an atomically-high steps that are separated by wide terraces. 

\begin{figure}[H]
\vspace{-0.2cm}
\centering
\includegraphics[width=4.5in]{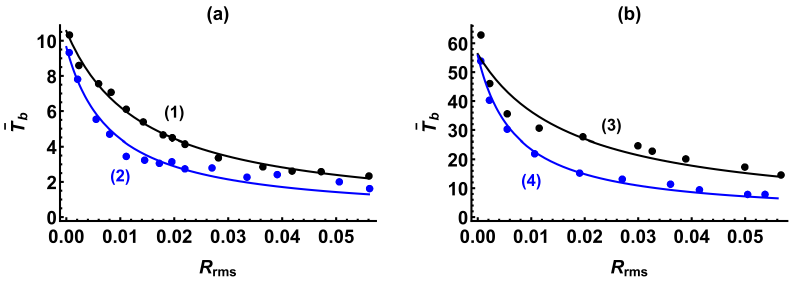}
\vspace{-0.15cm}
\caption{Average breakup time vs. $R_{rms}$ at (a) $E=5$ and (b) $E=0$. 
Black circles: $\bar N_\lambda= 3.5$; blue circles: $\bar N_\lambda= 4.5$. 
Parameters of the fitting inverse logarithmic curves are: (1): $p_1=0.15$, $p_2=1.01$; 
(2): $p_1=0.08$, $p_2=1.01$; (3): $p_1=1.00$, $p_2=1.02$; (4): $p_1=0.40$, $p_2=1.01$.
}
\label{BreakTime_vs_RMS_A=5_A=0}
\end{figure}

\begin{figure}[H]
\vspace{-0.2cm}
\centering
\includegraphics[width=4.5in]{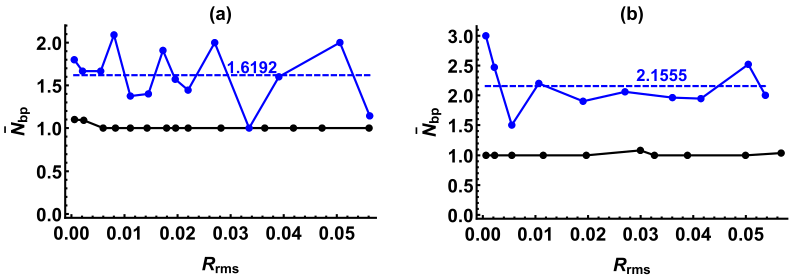}
\vspace{-0.15cm}
\caption{Average number of breakup points vs. $R_{rms}$ at (a) $E=5$ and (b) $E=0$. 
Black circles: $\bar N_\lambda= 3.5$; blue circles: $\bar N_\lambda= 4.5$. 
Solid curves are only the guides for the eye. Dashed lines and the numbers above them mark the mean values.
}
\label{NumberOFBreakUPS_vs_RMS_A=5_A=0}
\end{figure}

\begin{figure}[H]
\vspace{-0.2cm}
\centering
\includegraphics[width=4.5in]{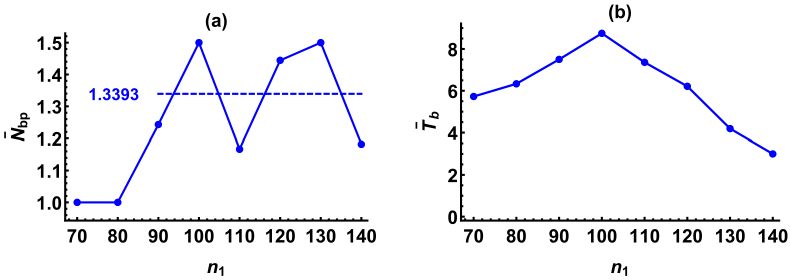}
\vspace{-0.15cm}
\caption{Average number of breakup points (a) and the average breakup time (b) vs. $n_1$. $E=5$, 
$R_{rms}=0.01$. Solid curves are only the guides for the eye. Dashed line in (a) and the number left of the line mark the mean value on the interval $90\le n_1\le 140$.
}
\label{NumberOFBreakUPS_BreakTime_vs_n1_A=5}
\end{figure}

A series of computations with varying $R_{rms}$ and $(n_1,n_2)$ pairs were conducted, with the goal of gaining the insight into how these parameters affect the breakup time and the number of 
breakups of a nanowire. For a measure of the axial roughness we adopt the number of 
``wavelengths"  of the initial surface profile, $N_\lambda$, per $\lambda_{max}$, averaged over all twenty segments comprising the computational domain. 
For the cases in Figures \ref{Fig_shapes_t=0}(a,c) these measures are $\bar N_\lambda= 3.5$ and $\bar N_\lambda=6.5$, respectively. The computational results
that are compared below were obtained mostly at $\bar N_\lambda= 3.5$ and $\bar N_\lambda= 4.5$; the latter value corresponds to $(n_1,n_2)=(90,110)$. 
\emph{To obtain each 
data point in Figures \ref{BreakTime_vs_RMS_A=5_A=0}-\ref{NumberOFBreakUPS_BreakTime_vs_n1_A=5}, we computed fifteen realizations of the 
initial condition (\ref{perturb}), and then averaged the outputs.} The standard deviation for these data sets ranges from 0.3 to 0.9.

\begin{figure}[H]
\vspace{-0.2cm}
\centering
\includegraphics[width=3.5in]{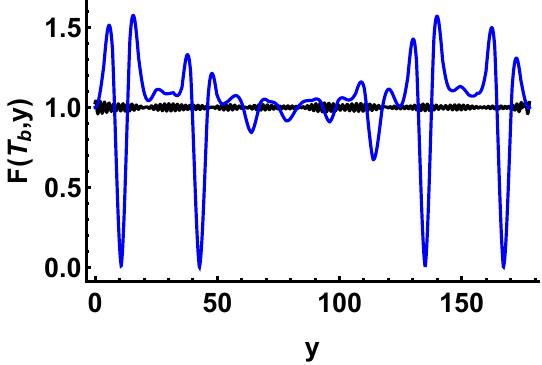}
\vspace{-0.15cm}
\caption{Blue curve: example profile of a nanowire surface at the breakup time. Black curve: initial rough surface.
}
\label{FinalShape}
\end{figure}

Computations were done using the Method of Lines framework in Mathematica\,\textsuperscript{\tiny\textregistered}. 
Periodic boundary conditions for Eq. (\ref{Vn}) were imposed at $y=0,\ \ell$. 
Partial derivatives in $y$ were discretized with a pseudospectral accuracy, the time integration was done using the variable-order backward difference method. 
The code is calibrated so that the computation terminates once the radius falls below 0.01 locally, i.e. at certain location along the wire axis $y$. 
Such point is the wire breakup point; one or more breakup points may occur simultaneously. In other words, a computation terminates when the local radius 
falls below 0.01 at one or more location along the wire axis. The time of the computation termination is taken as the wire breakup time.
Fig. \ref{FinalShape} shows, again for illustration only, one of the surface profiles at the breakup time $T_b$. This particular simulation resulted in 
four breakup points, thus in five cylindrical segments.

Figures \ref{BreakTime_vs_RMS_A=5_A=0}(a,b)  show the average wire breakup time $\bar T_b$ vs. $R_{rms}$. One observes that $\bar T_b$ is significantly 
smaller when the electromigration is operative. Increased axial roughness also contributes to a shorter breakup time. 
Data points in these Figures are nicely fitted with the inverse-logarithmic function 
$\bar T_b=p_1/\ln{\left(R_{\mbox{rms}}+p_2\right)}$, where $p_1$ and $p_2$ are the parameters. 
Their values are listed in the caption to Fig. \ref{BreakTime_vs_RMS_A=5_A=0}.  

Figures \ref{NumberOFBreakUPS_vs_RMS_A=5_A=0}(a,b)  show the average number of breakup points $\bar N_{bp}$ along a wire length $\ell$, vs. $R_{rms}$. At the smallest value of the axial roughness,
$\bar N_\lambda=3.5$, a wire breaks into only two cylindrical segments, irrespective of a value of the electromigration strength parameter $E$.
However, when the axial roughness is increased to $\bar N_\lambda=4.5$, the median number of break-up points increases to 1.62 and 2.15 at $E=5$ and $E=0$, respectively.
The median increase in the number of breakup points is 50\% higher when electromigration is not operative. 
Note also that at $E=0$ the average number of breakup points is very close to two, except at a small and large $R_{rms}$. The marked difference in $\bar N_{bp}$ values 
for two levels of the axial roughness is attributed to faster coarsening at larger $\bar N_\lambda$, i.e. when the characteristic axial 
roughness scale of the initial profile is smaller. Recall that for a deterministic, i.e. not randomized, 
initial condition such as a cosine curve profile with a wavenumber $k_{max}$, the number of breakup points would be twenty for the chosen wire 
length $\ell=20\lambda_{max}$. Thus faster coarsening creates the surface profile where the spatial variation is closer to $\lambda_{max}$, 
which results in more breakup points. 

Finally, in Figures \ref{NumberOFBreakUPS_BreakTime_vs_n1_A=5}(a,b) the average number of breakup points and the average breakup time are plotted vs. the axial roughness at
fixed $R_{rms}$ (only at $E=5$). Here $n_1$ is used as the measure of the axial roughness, since $\bar N_\lambda$ is difficult to calculate with high precision; 
for instance, $n_1=70$ corresponds to the pair
$\left(n_1,n_2\right)=(70,80)$,  $n_1=80$ corresponds to the pair $\left(n_1,n_2\right)=(80,90)$, etc. $(n_2-n_1=10.)$ In terms of $\bar N_\lambda$ values, 
$n_1=70$ corresponds to $\bar N_\lambda\approx 3.25$ and $n_1=140$ corresponds to $\bar N_\lambda\approx 7.2$. It is seen that multiple 
breakup points emerge when $\bar N_\lambda$ exceeds 3.4, corresponding to $\left(n_1,n_2\right)=(90,100)$. As $\bar N_\lambda$ increases, $\bar N_{bp}$ oscillates 
about the mean value 1.34 with a small amplitude 0.16.  Average breakup time initially slightly increases as $\bar N_\lambda$ increases and then sharply falls.

To summarize, computations in this letter show that the electromigrated nanowire breaks up into a cylindrical segments significantly faster.
A nanowire surface that is more rough in the radial or axial direction also results in a shorter breakup time, but the breakup time depends 
on the axial roughness non-monotonously. Another effect of a larger axial roughness is the increase in the number of cylindrical segments 
emerging after the breakup. 

The model in Ref. \cite{MyWire1} and in this paper allows for inclusion of a stress field, in a manner similar to Ref. \cite{KDMV}. 
Stress is expected to contribute to the instability of a wire surface and thus decrease the breakup time and likely increase the number of a breakup points. 
On the contrary, non-cylindrical wires, i.e. wires that have a polygonal cross-section, are more stable, since the corners on the wire surface need to get smoothed 
by surface diffusion before a breakup could occur. Analyses of these difficult problems require a significant investment of time and effort that is not presently possible, 
and thus they are deferred to future work.

\end{document}